\documentclass[aps,pra,twocolumn]{revtex4}
\usepackage{graphicx}
\usepackage{bm}
\usepackage{color}

\usepackage[applemac]{inputenc}

\begin{document}
\preprint{}
\title{Nonclassical joint distributions and Bell measurements}
\author{Elisa Masa, Laura Ares, and Alfredo Luis}
\email{alluis@fis.ucm.es}
\homepage{http://www.ucm.es/info/gioq}
\affiliation{Departamento de \'{O}ptica, Facultad de Ciencias
F\'{\i}sicas, Universidad Complutense, 28040 Madrid, Spain}
\date{\today}

\begin{abstract}
Derivation and experimental violation of Bell-like inequalities involve the measurement of incompatible observables. Simple complementarity forbids the existence of such joint probability distribution. Moreover, the measurement of incompatible observables requires different experimental procedures, which no necessarily must share a common joint statistics. In this work, we avoid these difficulties by proposing a joint simultaneous measurement. We can obtain the exact individual statistics of all the observables involved in the Bell inequalities after a suitable data inversion. A lack of positivity or any other pathology of the so retrieved joint distribution is then a signature of nonclassical behavior.
\end{abstract}

\pacs{42.25.-p Wave optics 
42.50.Dv	Quantum state engineering and measurements 
03.65.Ud	Entanglement and quantum nonlocality  
003.65.Ta	Foundations of quantum mechanics; measurement theory,  
42.50.Xa  Optical tests of quantum theory}

\maketitle

\section{Introduction}

Bell-like tests provide a powerful tool for testing, revealing and developing fundamental concepts at the very heart of quantum physics, involving quite basic ideas such as entanglement, reality, nonlocality, and even free will \cite{LB90,JB64,WW}. Violations of the Bell inequalities provide a very valuable insight in our knowledge of nature.  

\bigskip

A key point of Bell inequalities is that their derivation  and experimental violation involve the measurement and joint statistics of incompatible observables. This is a rather interesting point since it has been shown that Bell inequalities can be derived just from the simple hypothesis of the existence of a joint probability distribution for the observables involved, without resorting to realism or nonlocality \cite{AF82,AR15,BKO16,CH74,MA84,MC}. Therefore, from the very beginning, complementarity forbids the existence of such joint probability distribution, so the violation of Bell inequalities might not be so surprising after all. Alternatively, the measurement of incompatible observables requires different experimental procedures, that no necessarily must share a common joint statistics. This is a kind of practical contextuality, so to speak  \cite{MA84,MC,AK00,HP04,AM08,TN11,AK14,JCh17,NV62}. Thus, this complementarity issue really obscures the real meaning and consequences of the Bell inequalities.

\bigskip

In this work we propose an alternative approach that may avoid these difficulties. We propose that the statistics of all the observables involved in the Bell test are to be obtained from a single joint measurement of several compatible observables measured in one and the same experimental arrangement. These compatible observables can be regarded as fuzzy or noisy counterparts of the Bell-test observables. The key point is to design the  joint measurement such that its statistics provides the exact statistics of all the observables involved in the Bell inequalities after a suitable data inversion \cite{MM89,WMM02,WMM14,PB87, AL16,LM17,GBAL18}. 

The classical-physics version of such data inversion always provides the actual  joint probability distribution of the corresponding Bell-test observables, that always exists in classical physics. However, this is not the case in quantum mechanics since some of them are incompatible observables. In this work we show that lack of positivity or any other pathology of the so retrieved joint distribution is directly linked to the violation of the Bell inequalities \cite{AF82}. 

The point is that in this joint-measurement approach the final conclusions are no longer obscured nor diminished by practical contextuality problems. In this approach all observables are compatible and are measured in one and the same experimental arrangement. But nevertheless we show that they contain all the information required to test the Bell inequalities as required by the standard approach. 

\section{\label{sec:method}Method}
\subsection{\label{ssec:MS} System and joint measurement } 
As usual, our system will be made of two subsystems $A$ and $B$, both described by a two-dimensional Hilbert space, which may be the case for example of the polarization of two photons in two distinguishable field modes.

\bigskip

We consider arbitrary observables defined separately in each subsystem $A,B$. These are  $X,Y$ in the subsystem $A$, and $U,V$ in the subsystem $B$. All them will be described by positive operator-valued measures (POVMs) of the form 
\begin{equation}
\label{DW}
\Delta_W (w) = \frac{1}{2} \left  ( \sigma_0 +  w {\bm S}_W \cdot {\bm \sigma}  \right )   , 
\end{equation}
for $W=X,Y,U,V$, with $w=x,y,u,v=\pm 1$  the two only values allowed for each observable, being $\bm{\sigma}$ the Pauli matrices, and $\sigma_0$ is the identity. ${\bm S}_W$ are real vectors with $|{\bm S}_W| \leq 1$, so that $\Delta_W^\dagger (w) = \Delta_W (w)$ and $\Delta_W (w) >0$. 
These are dichotomic observables that are essentially spin or polarization measurements performed along a suitable spatial direction which is specified by ${\bm S}_W$.

\bigskip

For each subsystem  $A,B$ we consider a joint measurement of the corresponding two observables, described by the most general POVMs for dichotomic observables in a two-dimensional space
\begin{eqnarray}
\label{tAB}
& \tilde{\Delta}_A (x,y) = \frac{1}{4} \left  [ \sigma_0 +  \tilde{\bm{S}}_A (x,y) \cdot  \bm{\sigma}  \right ] , & \nonumber \\
 & & \\
 & \tilde{\Delta}_B (u,v) = \frac{1}{4} \left  [ \sigma_0 +   \tilde{\bm{S}}_B (u,v) \cdot  \bm{\sigma}  \right ]   ,  & \nonumber 
\end{eqnarray}
with
\begin{eqnarray}
\label{tS}
& \tilde{\bm{S}}_A (x,y) = x \gamma_X \bm{S}_X + y \gamma_Y \bm{S}_Y+ x y \gamma_{XY} \bm{S}_{XY},  & \nonumber  \\
 & & \\
 & \tilde{\bm{S}}_B (u,v) = u \gamma_U \bm{S}_U + v \gamma_V \bm{S}_V + u v \gamma_{UV} \bm{S}_{UV}, \nonumber
\end{eqnarray}
where $\gamma_{X,Y,U,V}$ are real factors expressing the accuracy in the observation of each observable, and $\gamma_{XY,UV}$ additional factors not related to the individual statistics but with information provided by the measurement beyond $X,Y$ and $U,V$. In any case, the $\gamma$ factor must be always chosen to ensure that $| \tilde{\bm{S}}_A (x,y) | \leq 1$, $| \tilde{\bm{S}}_B (u,v) | \leq 1$  so that $\tilde{\Delta}^\dagger_{A,B} = \tilde{\Delta}_{A,B}$ and $\tilde{\Delta}_{A,B} >0$. 

\bigskip

The complete measurement is then described by the POVM product
\begin{equation}
\label{cm}
\tilde{\Delta}_A \left ( x, y \right ) \otimes \tilde{\Delta}_B \left (u, v \right )  ,
\end{equation}
leading to a jointly observed statistics
\begin{equation}
\label{ptilde}
\tilde{p} ( x,y, u,v ) = \mathrm{tr} \left [ \rho \tilde{\Delta}_A \left ( x, y \right ) \otimes \tilde{\Delta}_B \left (u, v \right ) \right ] ,
\end{equation}
where $\rho$ is the density matrix of the complete system. These measurements are slightly more sophisticated than a simple Stern-Gerlach or a polarization measurements, being characterized by three spatial directions instead of just one. Nevertheless, they still admit suitable simple experimental implementations, and, for example, the joint statistics (\ref{ptilde}) can be obtained in an eight-port homodyne detector \cite{LM17,NGW87, LP96}.

Note that in principle the POVMs $\tilde{\Delta}_A \left ( x, y \right )$, $\tilde{\Delta}_B \left (u, v \right )$ involve all the three Pauli matrices, even though we are just interested in two dichotomic observables in each subsystem. This is because we are interested in deriving conclusions in the most general scenario possible. In this spirit we do not specify yet any particular expression for the vectors $\bm{S}_{X,Y,U,V}$, $\bm{S}_{XY,UV}$, since any choice would be equally valid regarding the conclusions. Nevertheless, for the sake of clarity we present some particular examples in Sec. 5.

\subsection{\label{ssec:MI}Inversion procedure}
The POVMs  $ \tilde{\Delta}_A \left ( x, y \right )$ and $\tilde{\Delta}_B \left (u, v \right )$ are assumed to provide complete information about the corresponding observables $X$, $Y$ and $U$, $V$, respectively, so that the exact statistics of each $\Delta_W$ is contained in the corresponding marginal $\tilde{\Delta}_W$. For example, for $W=X$ we have the marginal POVM
\begin{equation}
\tilde{\Delta}_X (x) =   \sum_{y=\pm 1} \tilde{\Delta}_A \left ( x, y \right ) ,
\end{equation}
leading to 
\begin{equation}
\label{marg}
\tilde{\Delta}_X (x) = \frac{1}{2} \left  ( \sigma_0 +  x \gamma_X {\bm S}_X \cdot {\bm \sigma}  \right )   , 
\end{equation}
and equivalently for all the other POVMs.

\bigskip

Thus, we assume that for each $W$ there are state-independent functions $\mu_W (w,w^\prime) $ so that the exact $\Delta_W (w)$  in Eq. (\ref{DW}) is obtained from the marginal $\tilde{\Delta}_W (w)$ in Eq. (\ref{marg}) as 
\begin{equation}
\Delta_W (w) =   \sum_{w^\prime = \pm 1} \mu_W (w, w^\prime ) \tilde{\Delta}_W \left (w ^\prime \right ) .
\end{equation}
It can be easily seen that these functions are 
\begin{equation}
\mu_W (w,w^\prime) = \frac{1}{2} \left ( 1 + \frac{w w^\prime}{ \gamma_W}  \right ).
\end{equation}

\bigskip

Applying the inversion procedure to the complete measurement (\ref{cm}) we get the family of operators $\Delta_A \left ( x, y \right ) \otimes \Delta_B \left (u, v \right )$ where 
\begin{eqnarray}
& \Delta_A \left ( x, y \right ) =  \sum_{x^\prime, y^\prime=\pm 1} \mu_X (x, x^\prime ) \mu_Y (y, y^\prime) \tilde{\Delta}_A \left ( x^\prime , y^\prime \right ) , \nonumber \\
 & & \\
& \Delta_B \left ( u, v \right ) =  \sum_{u^\prime , v^\prime=\pm 1} \mu_U (u, u^\prime ) \mu_V (v, v^\prime ) \tilde{\Delta}_B \left ( u^\prime , v^\prime \right ) ,\nonumber 
\end{eqnarray}
leading to 
\begin{eqnarray}
\label{AB}
& \Delta_A (x,y) = \frac{1}{4} \left  [ \sigma_0 + \bm{S}_A (x,y) \cdot  \bm{\sigma}  \right ] , & \nonumber \\
 & & \\
 & \Delta_B (u,v) = \frac{1}{4} \left  [ \sigma_0 + \bm{S}_B (u,v) \cdot  \bm{\sigma}  \right ]   ,  & \nonumber 
\end{eqnarray}
with
\begin{eqnarray}
\label{tSAB}
\label{SAB}
& \bm{S}_A (x,y) = x  \bm{S}_X + y  \bm{S}_Y + x y \frac{\gamma_{XY}}{\gamma_X \gamma_Y} \bm{S}_{XY} , & \nonumber  \\
 & & \\
 & \bm{S}_B (u,v) = u \bm{S}_U + v  \bm{S}_V + u v \frac{\gamma_{UV}}{\gamma_U \gamma_V} \bm{S}_{UV}  . \nonumber
\end{eqnarray}
Finally the inferred joint distribution is obtained as
\begin{equation}
\label{ijd}
p ( x,y, u,v ) = \mathrm{tr} \left [ \rho \Delta_A \left ( x, y \right ) \otimes \Delta_B \left (u, v \right ) \right ] .
\end{equation}
We have to stress that in classical physics this inversion procedure always leads to a proper joint probability distribution \cite{AL16,LM17}.

\section{\label{sec:IAS}Inversion and statistics for an arbitrary state}

Now we apply the inversion procedure to a general case in order to analyze the possibles pathologies of the joint distribution in parallel with the main features of the Bell scenario.

\subsection{System state} 
For the purpose of obtaining the maximum generality, we describe the complete system state via the most general density matrix that can be always expressed as \cite{DGGA,BM15}
\begin{eqnarray}
\label{stategeneral}
&\rho =\frac{1}{4} ( \sigma_0 \otimes \sigma_0 +\\
&\sum\limits_{i=1}^3 a_{i} \sigma_i \otimes \sigma_0+ \sum\limits_{i=1}^3 b_i \sigma_0 \otimes \sigma_i + \sum\limits_{i=1}^3 c_{i} \sigma_i \otimes \sigma_i ) , \nonumber
\end{eqnarray}
for $i=1,2,3$, where $\sigma_i$ are the Pauli matrices and $\bm{a}=\{a_i\}, \bm{b}=\{b_i\}, \bm{c}=\{c_i\} $ are real vectors. %a and b with modulus les than or equal to one?

\subsection{Obtaining statistics}  Equivalent expressions hold
For the state (\ref{stategeneral}) we have that the observed joint statistics and the inferred distribution are: 
\begin{eqnarray}
&\tilde{p} ( x,y, u,v ) = \\ &\frac{1}{16} \left [1 + \bm{a} \cdot \bm{\tilde{S}}_A (x,y) +\bm{b} \cdot \bm{\tilde{S}}_B (u,v) + \sum\limits_{i=1}^3 c_i \tilde{S}_{A_i}\tilde{S}_{B_i}\right ] ,\nonumber
\end{eqnarray}
and  
\begin{eqnarray}
\label{pinf}
&p ( x,y, u,v ) = \\ &\frac{1}{16} \left [1 + \bm{a} \cdot \bm{S}_A(x,y) +\bm{b} \cdot \bm{S}_B (u,v)+ \sum\limits_{i=1}^3 c_i S_{A_i}S_{B_i}\right ] ,\nonumber
\end{eqnarray}
where $\bm{\tilde{S}}_A $, $\bm{\tilde{S}}_B$, $\bm{S}_A $ and $\bm{S}_B$ are in Eqs. (\ref{tS}) and (\ref{SAB}), while $S_{A_i}$ are the corresponding  vector components.

\bigskip

Let us stress that there are restrictions on the modulus of $\tilde{\bm{S}}_{A,B}$ since they determine the statistics of measured observables, and also restrictions on the modulus of all $\bm{S}_{X,Y,U,V}$ since they determine the true statistics of the corresponding observables. However, there are no such restrictions of  modulus for $\bm{S}_{A,B}$.  And this allows for the appearance of pathologies in the inferred statistics (\ref{pinf}) depending on the system state $\rho$. 

\bigskip

Before showing this let us examine  all marginals of this joint distribution $p(x,y,u,v)$ to properly complete the analysis. The one-observable marginals are by construction the exact ones:
\begin{equation}
\label{margWx}
p_{W} (w) =   \frac{1}{2} (1 + w \bm{a}\cdot\bm{S_W})
\end{equation}
for $W=X,Y$, $w=x,y$, and 
\begin{equation}
\label{margWu}
p_{W} (w) =   \frac{1}{2} (1 + w \bm{b}\cdot\bm{S_W})
\end{equation}
for $W=U,V$, $w=u,v$. 
\bigskip
The joint $A$ and $B$ marginals are:
\begin{eqnarray}
\label{margXY}
&p_{X,Y} (x,y) = \frac{1}{4}[1+\bm{a}\cdot\bm{S_A
}(x,y)], \nonumber \\ & &  \\
&p_{U,V} (x,y) = \frac{1}{4}[1+\bm{b}\cdot\bm{S_B
}(u,v)]. \nonumber
\end{eqnarray}
\bigskip
The cross two-observable marginals:
\begin{eqnarray}
\label{margXU}
&p_{X,U} (x,u) = \\
&\frac{1}{4} \left ( 1 + u \bm{b} \cdot \bm{S}_U + x \bm{a} \cdot \bm{S}_x + x u  \sum c_{i} S_{X_i} S_{U_i}\right),\nonumber 
\end{eqnarray}
and equivalently for $p_{X,V} $, $p_{Y,U} $, and $p_{Y,V}$. 
At difference with Eq. (\ref{margXY}) this is the joint statistics of two commuting observables $X$ and $U$, so it has no problem to be defined directly in terms of their common eigenvectors. Let us briefly show that actually Eq. (\ref{margXU}) is exactly the true joint distribution for $X$ and $U$, as a further proof of the validity of our method. To this end we recall that the POVM elements associated to $X$ and $U$ are, in their corresponding Hilbert spaces,
\begin{equation}
\Delta_X (x) =   \frac{1}{2} (\sigma_0 + x \bm{S}_X \cdot\bm{\sigma}) , \quad
\Delta_U (u) =   \frac{1}{2}  (\sigma_0 + u \bm{S}_U \cdot\bm{\sigma}) ,
\end{equation}
so that the joint statistics is 
\begin{equation}
p_{X,U} (x,u) = \mathrm{tr} \left [ \rho \Delta_X (x) \otimes \Delta_U (u) \right ] .
\end{equation}
For the state (\ref{stategeneral}) and using the good properties of Pauli matrices under the trace we readily get that this is exactly the same in Eq. (\ref{margXU}).

\bigskip

More interesting results arise for three-observable marginals such as
\begin{eqnarray}
\label{margXUV}
&p_{X,U,V} (x,u,v)= \\
&\frac{1}{8} \left [ 1 +  \bm{b} \cdot \bm{S}_B (u,v) + x  \bm{a} \cdot \bm{S}_X+ x \sum\limits_{i=1}^3 c_{i} S_{X_i} S_{B_i} (u, v )  \right ].\nonumber
\end{eqnarray}
Equivalent expressions hold for $p_{Y,U,V} $, $p_{X,Y,U} $, and $p_{X,Y,V} $. 
These are maybe the most interesting marginals for our purposes since it is the pathology of these distributions which is responsible for the violation of Bell inequalities, as we show next. This is because they are the simplest distributions containing both observables of subsystems A and B as well as two complementary observables such as $X$ and $Y$.

\section{Pathology and violation of Bell inequalities}
Let us show explicitly that violation of Bell-like inequalities implies that the retrieved joint distribution $p(x,y,u,v)$ takes negative values, in agreement with Fine's theorem \cite{AF82}. To this end we use Bell-like inequalities expressed directly in terms of probabilities \cite{CH74,AF82}. This is that the existence of a legitimate probability distribution $p(x,y,u,v)$ implies the satisfaction of four double inequalities of the form, for $x,u = \pm 1$:
\begin{eqnarray}
\label{Ineq}
& 0 \geq p_{X,U} (x,u) - p_{X,V} (x,v) + p_{Y,U} (y,u) + p_{Y,V} (y,v)& \nonumber \\
& - p_Y (y) - p_U (u)  \geq -1, &
\end{eqnarray} 
and three other equivalent relations obtained by suitably permuting the observables. 

We define $C$ as the nucleus of the Bell-like double inequality (\ref{Ineq}):
\begin{eqnarray}
\label{BCH}
&C = p_{X,U} (x,u) - p_{X,V} (x,v) +\\
&p_{Y,U} (y,u) + p_{Y,V} (y,v)- p_Y (y) - p_U (u) ,\nonumber
\end{eqnarray}
so that the inequalites are
\begin{equation}
\label{CC}
        0 \geq C \geq -1 .
\end{equation}

\bigskip

Now, starting from the expressions (\ref{margXU}) and (\ref{margXUV}) for the two- and three-observable marginals, it can be proved that
\begin{eqnarray}
\label{Sigma}
& \Sigma = p_{X,U,V} (x,u,-v) + p_{X,U,V} (-x,-u,v) &  \\
& + p_{Y,U,V} (y,u,v) + p_{Y,U,V} (-y,-u,-v)  \nonumber \\
&= C+1,\nonumber
\end{eqnarray}
that can be derived also using general  properties of the probability distributions.
Note that if there were a legitimate joint distribution then $1 \geq \Sigma \geq 0$. It must be nonnegative since it is a sum of probabilities and is less than or equal to one since, for all $W=X,Y$ and $w=x,y = \pm 1$ we have
\begin{equation}
p_{W,U,V} (w,u,v) \leq p_{U,V} (u,v) ,
\end{equation}
and therefore

\begin{eqnarray}
& \Sigma \leq  p_{U,V} (u,-v) + p_{U,V} (-u,v)  & \nonumber \\ & + p_{U,V} (u,v) + p_{U,V} (-u,-v) = 1 . & 
\end{eqnarray}

If we replace $C=\Sigma-1$ in Eq. (\ref{CC}),
the Bell-like inequalities are equivalent to:
\begin{equation}
\label{Ineq2}
    1 \geq \Sigma  \geq 0  .
\end{equation}
Therefore any of the violations of the Bell-like inequalities means either $\Sigma <0$ or $\Sigma >1$, which are incompatible with a legitimate joint distribution. 

Consequently, our scheme provides a nice and useful realization of the more abstract scenarios on the subject.

\section{Examples}
For the sake of clarity let us illustrate our approach with some typical examples of Bell tests. As physical system we consider the usual singlet state, that has the same form in any basis, 
\begin{equation}
    |\psi \rangle = \frac{1}{\sqrt{2}} \left ( | + \rangle_A | - \rangle_B -  | - \rangle_A | + \rangle_B \right ) ,
\end{equation}
where $ | \pm  \rangle_{A,B}$ are the eigenstates of a given Pauli matrix in the corresponding subsystem. In the density-matrix form in Eq. (\ref{stategeneral}) this is 
\begin{equation}
\rho =\frac{1}{4} \left ( \sigma_0 \otimes \sigma_0 - \sum\limits_{i=1}^3  \sigma_i \otimes \sigma_i \right ) . 
\end{equation}
Regarding the measurements invoked in the Bell test let us consider a very simple scenario with observables measured along the directions in some plane $XY$ specified by angles $\alpha$, $\alpha^\prime$, $\beta$ and $\beta^\prime$
\begin{equation}
    \bm{S}_X = \left (\cos \alpha ,\sin \alpha, 0 \right ),  \quad \bm{S}_Y = \left (\cos \alpha^\prime ,\sin \alpha^\prime , 0 \right ),
\end{equation}
and
\begin{equation}
    \bm{S}_U = \left (\cos \beta ,\sin \beta, 0 \right ),  \quad \bm{S}_Y = \left (\cos \beta^\prime ,\sin \beta^\prime , 0 \right ),
\end{equation}
with 
\begin{equation}
    \bm{S}_{XY} =  \bm{S}_{UV} = \bm{0} .
\end{equation}

Let us consider two simple options. In the first case 
\begin{equation}
    \alpha =0, \quad   \alpha^\prime = \frac{\pi}{2}, \quad   \beta = \frac{3 \pi}{4}, \quad \beta^\prime = -\frac{3 \pi}{4},  
\end{equation}
for which, by direct computation of the corresponding probabilities we get for $x=-y=-u=v=1$ that $C$ and $\Sigma$ in Eqs.(\ref{BCH}) and (\ref{Sigma}) are
\begin{equation}
C = -\frac{1}{2}-\frac{1}{\sqrt{2}} = -1.21, \quad \Sigma = \frac{1}{2}-\frac{1}{\sqrt{2}} = -0.21,
\end{equation}
violating Bell inequality with a retrieved joint distribution $p(x,y,u,v)$ incompatible with standard statistics. In particular $p(1-1,-1,1)= -0.026$.

As a second typical example let us consider   
\begin{equation}
    \alpha =0, \quad \alpha^\prime = 0.375 \pi, \quad   \beta = 2 \alpha^\prime, \quad \beta^\prime = 3 \alpha^\prime,  
\end{equation}
for which, by direct computation of the corresponding probabilities we get for the same $x=-y=-u=v=1$ that 
\begin{equation}
C  = -1.18, \quad \Sigma = -0.18,
\end{equation}
violating Bell inequality with a $p(x,y,u,v)$ incompatible with standard statistics, in particular $p(-1,-1,1,1)= -0.06$.  

\section{Conclusions}
We have presented an alternative approach to Bell-like scenarios with many fruitful possibilities. It provides a suitable arena to investigate basic quantum features such as complementarity, entanglement, nonclassicality within one and the same framework, avoiding previous assumptions that may obscure the proper elucidation of these basic features. 

\section*{ACKNOWLEDGMENTS}
L. A. and A. L. acknowledge financial support from Spanish Ministerio de Econom\'ia y Competitividad Project No. FIS2016-75199-P.
L. A. acknowledges financial support from European Social Fund and the Spanish Ministerio de Ciencia Innovaci\'{o}n y Universidades, Contract Grant No. BES-2017-081942.

\end{document}